\documentclass[graybox]{svmult}

\usepackage{amssymb}

\usepackage{mathptmx}       
\usepackage{helvet}         
\usepackage{courier}        
\usepackage{type1cm}        
\usepackage{makeidx}         
\usepackage{graphicx}        
\usepackage{multicol}        
\usepackage[bottom]{footmisc}

\makeindex             

\newcommand{\pc}{\Phi_\mathrm{c}}
\newcommand{\mc}{M_\mathrm{c}}

\begin{document}

\title*{Self-gravitating Newtonian disks revisited}
\author{Patryk Mach, Edward Malec and Walter Simon}
\institute{Patryk Mach, Edward Malec \at M. Smoluchowski Institute of Physics, Jagiellonian University, Reymonta 4, 30-059 Krak\'{o}w, Poland \email{patryk.mach@uj.edu.pl, malec@th.if.uj.edu.pl}
\and Walter Simon \at Gravitational Physics, Faculty of Physics, Vienna University, Boltzmanngasse 5, 1090 Vienna, Austria \email{walter.simon@univie.ac.at}}
%
%
\maketitle

\abstract{Recent analytic results concerning stationary, self-gravitating fluids in Newtonian theory are discussed. We give a theorem that forbids infinitely extended fluids, depending on the assumed equation of state and the rotation law. This part extends previous results that have been obtained for static configurations. The second part discusses a Sobolev bound on the mass of the fluid and a rigorous Jeans-type inequality that is valid in the stationary case.}

\section{Introduction}

In a series of papers \cite{mach_malec_2011, mach_2012, mach_simon} we deal with the properties of self-gravitating, stationary configurations of barotropic fluids. In the Newtonian theory, the underlying set of equations consist of the continuity equation
\begin{equation}
\label{cont}
\nabla \cdot (\rho \mathbf U) = 0,
\end{equation}
the Euler equation
\begin{equation}
\label{euler}
\nabla \cdot (\rho \mathbf U \otimes \mathbf U) + \nabla p + \rho \nabla \Phi = 0,
\end{equation}
and the Poisson equation for the gravitational potential
\begin{equation}
\label{poisson}
\Delta \Phi = 4 \pi G \rho.
\end{equation}
Here $\rho$ denotes the mass density, $p$ is the pressure, $\mathbf U$ the velocity of the fluid, and $\Phi$ the gravitational potential. The above system is closed by the barotropic equation of state $p = p(\rho)$ and the rotation law.

From the physical point of view solutions of Eqs.~(\ref{cont}--\ref{poisson}) serve as simple models of rotating stars, or self-gravitating accretion disks (note that a disk-like or toroidal shape of the fluid region is allowed). The key mathematical questions include the existence of solutions, their uniqueness, possible parametrization and stability.

In the following we will restrict ourselves to configurations for which the term $(\mathbf U \cdot \nabla) \mathbf U$ can be expressed as a gradient of a potential, i.e., $(\mathbf U \cdot \nabla) \mathbf U = \nabla \pc$. In this case the Euler equations can be integrated, yielding
\begin{equation}
\label{int_euler}
h + \pc + \Phi = C,
\end{equation}
where $h$ denotes the specific enthalpy $dh = dp/\rho$, and $C$ is an integration constant. Equation (\ref{int_euler}) is only valid in the closure of a region where $\rho \neq 0$. In what follows, this region will be denoted by $\Omega$.

\section{Finite or infinite?}

In this section we formulate simple criteria on the equation of state $p = p(\rho)$ and the rotation law $\pc = \pc(\mathbf x)$ that guarantee finite spatial extent of a stationary configuration of the rotating fluid. The detailed discussion can be found in \cite{mach_simon}. It extends previous results that have been obtained for the static case, both in General Relativity and in Newtonian Theory \cite{simon, heinzle, uggla}.

The argument is based on the suitable form of the virial theorem. Consider a vector field $\mathbf w = \left( (\mathbf x \cdot \nabla) \Phi + \frac{1}{2} \Phi \right) \nabla \Phi - \frac{1}{2} | \nabla \Phi|^2 \mathbf x + 4 \pi G p \mathbf x$. A simple calculation making use of Eqs.~(\ref{cont}, \ref{euler}, \ref{poisson}) shows that $\nabla \cdot \mathbf w = 4 \pi G \left( \frac{1}{2} \rho \Phi - \rho \mathbf x \cdot \nabla \pc + 3 p \right)$. Integrating the above equation over $\mathbb R^3$ one obtains
\begin{equation}
\label{virial}
\frac{1}{2} \int_{\mathbb R^3} d^3 x \rho \Phi -  \int_{\mathbb R^3}d^3 x \rho \mathbf{x} \cdot \nabla \pc + 3 \int_{\mathbb R^3} d^3 x p = 0,
\end{equation}
provided that the surface integral of $\mathbf w$ vanishes at infinity. This is assured by requiring that $\rho \in W^{0,2}_{-3-\epsilon}$, $0 < \epsilon < 1$. Then $\Phi \in W^{2,2}_\mathrm{loc}$, and $\Phi - M/\sigma \in W^{2,2}_{-1-\epsilon}$. Here we are using weighted Sobolev spaces $W^{k,p}_\delta$ ($1 \le p \in \mathbb R$, $\delta \in \mathbb R$, $k \in \mathbb N_0$) based on weighted Lebesgue norms
\[ \Vert u \Vert_{k,p,\delta} = \sum_{0 \le | \alpha | \le k} \Vert D^\alpha u \Vert_{p,\delta - |\alpha|}, \;\;\; \Vert u \Vert_{p,\delta} = \left( \int_{\mathbb R^3} d^3 x |u|^p \sigma^{-\delta p - 3} \right)^{1/p} \]
with $\sigma = (1 + |\mathbf x|)^{1/2}$. The total mass of the fluid $M = \int_{\mathbb R^3} d^3 x \rho$ is assumed to be finite. We also require that $p \in W^{1,1}_{-4-\epsilon}$.

The integrated Euler equation (\ref{int_euler}) and Eq.~(\ref{virial}) yield $MC = \int_{\mathbb R^3} d^3 x \left(F + 2 \rho D \right)$, where $F = \rho h - 6 p$, $D = \mathbf x \cdot \nabla \pc + \frac{1}{2} \pc$. The key observation, following from Eq.~(\ref{int_euler}), is that an infinite configuration with a finite mass requires $C =0$. Thus if $F > 0$ and $D > 0$, or $F < 0$ and $D < 0$, the fluid must be finite.

We would like to point out the role of the assumed rotation law $D = D(\mathbf x)$. Former results on the finiteness of configurations were basically amendments to existence theorems \cite{auchmuty}, and finiteness was assured by imposing stringent conditions on the equation of state only.

The limiting-case conditions $F \equiv 0$ and $D \equiv 0$ lead to a polytropic equation of state $p = K \rho^{1+1/n}$ with $n = 5$, and $\pc = z^{-1/2} \tau(x/z, y/z)$, where $\tau$ is an arbitrary function, and $(x,y,z)$ denote Cartesian coordinates. Remarkably, the resulting equation for $h$ is invariant under the scaling transformation $h(\mathbf x) \mapsto h ( \mathbf x/\lambda )/\sqrt{\lambda}$, $\lambda \in \mathbb R_+$.

Another result can be obtained for axially symmetric systems. Let $(r,\phi, z)$ denote the cylindrical coordinates. A classic theorem due to Poincar\'{e} and Wavre \cite{tassoul} states that for barotropic fluids with $\mathbf U = \omega(r,z) \partial_\phi$ one has in fact $\omega = \omega(r)$, and $\pc = - \int^r dr^\prime r^\prime \omega^2 (r^\prime)$. In this case the fluid cannot extend to infinity in the $z$ direction, unless it is static. This follows by contradiction from Eq.~(\ref{int_euler}).

Systems with a central point mass $\mc$, that resemble disk-like configurations around compact objects, should be discussed separately. The suitable form of the virial theorem for such systems was formulated in \cite{mach_2012}, and the reasoning concerning finite extent of stationary configurations was done in \cite{kepl2}.

\section{Mass estimates}

In the following, we give a strict derivation of a new mass estimate valid for a class of stationary configurations of perfect fluids. The discussion is based on a paper by Mach and Malec \cite{mach_malec_2011}. We specialize to polytropic equations of state $p = K \rho^{1+1/n}$, where $K$ and $n$ are constant, and assume that the fluid is finite. From Eqs.~(\ref{cont}) and (\ref{euler}) one obtains
\begin{equation}
\label{bbb}
\Delta h = - A h^n - \Delta \pc,
\end{equation}
where $A = 4 \pi G/(K(1+n))^n$. Assume now that $\Delta \pc \leq 0$. Multiplying Eq.~(\ref{bbb}) by $h$ and integrating over $\Omega$ we get
\[ - \int_\Omega d^3 x h \Delta h = \int_\Omega d^3 x |\nabla h|^2 = A \int_\Omega d^3x h^{n+1} + \int_\Omega d^3x h \Delta \pc
 \le  A \int_\Omega d^3x h^{n+1}, \]
because $h = 0$ on $\partial \Omega$. The last integral in the above expression can be estimated making use of the H\"{o}lder inequality. For $n > 1$, we have
\[ \int_\Omega d^3 x h^{n+1} = \int_\Omega d^3 x h^{n-1}h^2 \le \Vert h^{n-1} \Vert_{L^{3/2}(\Omega)} \Vert h^2 \Vert_{L^3(\Omega)}. \]
The Sobolev inequality yields $\Vert h \Vert_{L^6(\Omega)} \le C(3,2) \Vert \nabla h \Vert_{L^2(\Omega)}$, where the constant $C(3,2) = 4^{1/3}/(\sqrt{3} \pi^{2/3})$ is a universal number in $\mathbb R^3$. This gives
\[ \Vert h^{n-1} \Vert_{L^{3/2}(\Omega)} = \left( \int d^3 x h^{3(n-1)/2} \right)^{2/3} \ge \frac{1}{A C^2(3,2)}. \]

The remaining steps are simple. Introducing the mass of the fluid $M = \int_\Omega d^3 x \rho = A/(4 \pi G) \int_\Omega d^3 x h^n$ we get
\begin{equation}
\label{ineq1}
M > \left ( 4 \pi G \sqrt{A} C^3(3,2) h_\mathrm{max}^{(n-3)/2} \right)^{-1}
\end{equation}
for $n \ge 3$. Here $h_\mathrm{max}$ denotes the maximum value of the specific enthalpy. For the ideal gas the temperature is given by $T = p \mu m_\mathrm{p}/(\rho k_\mathrm{B})$, where $\mu$ is the mean molecular weight, $m_\mathrm{p}$ denotes the proton mass, and $k_\mathrm{B}$ is the Boltzmann constant. In this case the inequality (\ref{ineq1}) can be written as
\[  M > \frac{3 \sqrt{3 \pi}}{32} \left( \frac{(1+n) k_\mathrm{B}}{G \mu m_\mathrm{p}} \right)^{3/2} \frac{T_\mathrm{max}^{3/2}}{\sqrt{\rho_\mathrm{max}}}.  \]

An interesting corollary follows from the above discussion for $n = 3$. Let $\bar \rho$ and $\bar T$ denote volume averaged mass density and temperature, respectively. A simple calculation involving H\"{o}lder's inequality yields $\bar T \le (K \mu m_\mathrm{p}/k_\mathrm{B}) \bar \rho^{1/n}$. The mass estimate can be now written as
\[ M > \frac{3\sqrt{3 \pi}}{4} \left( \frac{k_\mathrm{B}}{G \mu m_\mathrm{p}} \right)^{3/2} \frac{\bar T^{3/2}}{\sqrt{\bar \rho}}, \]
i.e., in a form of a Jeans inequality for a bound system.

\end{document}